\documentclass[superscriptaddress,twocolumn,nofootinbib]{revtex4}
\usepackage{graphicx}
\usepackage{color}
\usepackage{enumerate}
\usepackage{amssymb}
\usepackage{bbm}
\usepackage{dsfont}
\def\be{\begin{equation}} 
\def\ee{\end{equation}} 

\def\ve{v_{\rm eff}}
\def\unit{\mathds{1}}
\renewcommand{\vec}[1]{\mbox{\boldmath $#1$}}
\def\vH{\mbox{\bf H}}
\def\vV{\mbox{\bf V}}
\def\vG{\mbox{\bf G}}
\def\vA{\mbox{\bf A}}

\begin{document}

\title{
Least action and the maximum-coupling approximations \\ 
in the theory of spontaneous fission
}

\author{K. Hagino}
\affiliation{ 
Department of Physics, Kyoto University, Kyoto 606-8502,  Japan} 

\author{G.F. Bertsch}
\affiliation{ 
Department of Physics and Institute of Nuclear Theory, Box 351560, 
University of Washington, Seattle, Washington 98915, USA}


\begin{abstract}
We investigate the dynamics of spontaneous fission in a
configuration-interaction (CI) approach.  In that formalism
the decay rate is governed by an effective  interaction
coupling the ground-state configuration and a fission doorway
configuration, with the interaction strength determined by inverting
a high-dimensioned CI Hamiltonian matrix that may have a
block-tridiagonal structure.  It is shown that the decay rate
decreases exponentially with the number of blocks at a rate
determined by the largest eigenvalue of a matrix in the block 
space for Hamiltonians with identical off-diagonal blocks.
The theory is greatly
simplified by approximations  
similar in spirit to the adiabatic and the least-action approximations
in continuum representations.  Here each block is replaced by a 
single matrix element.   While the adiabatic reduction 
underestimates the coupling,  a reduction based on a maximum-coupling
approximation works well in a schematic CI model.  
\end{abstract}


\maketitle

\section{Introduction}

The theory of spontaneous fission 
is a challenging subject of multidimensional quantum tunneling.  The theory
is usually formulated by defining one or more collective coordinates
in a constrained mean-field theory and then mapping the Hamiltonian or energy functional onto 
a Schr\"odinger equation in
those coordinates.  See for example Refs. \cite{Moller2009,
Shunck2016,D19,whitepaper2020}. 
Once a tunneling
path through the collective space is determined,
the decay rate is calculated from the 
collective potential and inertia using the WKB 
formula. 

The configuration 
interaction (CI) approach offers a completely different framework
for calculating the large-amplitude dynamics needed in 
fission theory.  Instead of invoking collective coordinates to describe
the dynamics, the theory is based on the Hamiltonian interactions
between configurations in a discrete basis.   Not all of the tools for 
carrying out realistic calculations are presently in place, but
several aspects have been demonstrated.  In particular, one need not 
rely entirely on collective coordinates to construct the needed
configuration spaces \cite{BM16}.  Also, it is feasible to estimate
the decay widths of doorway states into fission channels
with available calculational tools \cite{BR19}. 
There are at least three advantages of the CI approach over the conventional 
approaches that can be mentioned.  First, 
the overcompleteness problem inherent 
in the generator coordinate method (GCM) can be mitigated.  Second,
couplings to intrinsic excitations can be 
incorporated relatively easily. Finally, the Hamiltonian formulation 
is particularly suited for calculating fission cross sections in the
$K$-matrix reaction theory \cite{Kawano15,Bertsch20}. 

We have previously explored  a schematic  
model  based on the discrete basis approach, with application to
induced fission \cite{Bertsch20} 
and  to spontaneous fission \cite{HB20}. The Hamiltonian
in the schematic model is simple enough 
to be fully solvable, so that it can serve as a test of existing approximations. 
One of the important findings in Ref. \cite{HB20} is that the adiabatic approximation, 
which is deeply embedded in the theory of spontaneous fission, may significantly 
underestimate the decay width. This has already been shown in realistic
calculations in the WKB framework comparing the adiabatic treatment with
the more sophisticated least-action approach \cite{Robledo14,Robledo18,Moretto74}.  
Clearly there
is a need to understand the accuracy of the approximations inherent in
the different approaches.  

In this paper we propose a new approximation scheme within the CI framework,
called here the 
``maximum-coupling approximation".  It has a close
resemblance to the least-action approximation in the WKB formalism.
Since the model is exactly solvable numerically, its accuracy can
be tested.  It was shown in the pioneering study of 
Moretto and Babinet \cite{Moretto74} that treating the pairing field as
a dynamic variable strongly affects the calculated action in the tunneling
region.  This conclusion was recently re-affirmed by realistic calculations 
of the action using the generator coordinate method GCM \cite{Robledo14,Robledo18}.
The same idea can be also
implemented in the discrete-basis approach.  This is
done by increasing the pairing interaction to construct intermediate 
configurations at each shape parameter. 
As we will show, this approximation 
reproduces quite well the decay width that is obtained with the full 
Hamiltonian. 

The paper is organized as follows.  In Sec.  II, we introduce the
general framework for calculating transport in a CI basis, as well as the
approximations and reductions made for dealing with large spaces.
In Sec. III we apply the theory to the schematic
model proposed in our earlier publications,  comparing exact numerical
calculations to approximate
treatments including the maximum-coupling approximation. Finally
in Sec. IV we discuss the relationship to the least-action 
approach.  

\section{Discrete-basis approach}

We assume that a basis of many-body states has been constructed \cite{BY19,BYR18,BYR19}
to calculate the decay from a ground-state configuration\footnote{
The term ``ground-state" should be qualified:  we do not mean 
the true eigenstate but only the state representing it in the
many-body configuration space.} through a doorway configuration for a particular
decay channel.  Because there is a large-amplitude reorganization
of orbitals and occupation factors in the transition,
many intermediate configurations must be included in the space.
 
\subsection{Formulation}

The general form of the Hamiltonian matrix is 
\be
\vH = \left(
\matrix{
E_{g} & \vec{v}_g^T & 0 \cr
\vec{v}_g  & \vH_b & \vec{v}_d  \cr
0 & \vec{v}_d^T & E_d \cr}
\right),
\label{H}
\ee
in a notation using Roman boldface type font for matrices.
$E_g$ and $E_d$ are the energies of the ground-state and
doorway configurations, $\vH_b$ (for ``barrier") is the Hamiltonian  matrix of
the intermediate configurations, and $\vec{v}_g$ and $\vec{v}_d$ are vector
arrays of the matrix elements coupling $\vH_b$ 
to the two end-point configurations.  The energy of the doorway
configuration has an imaginary part $\Gamma_d/2$, which may be 
evaluated with the Fermi golden rule as has been argued in Refs. \cite{BR19,BY19}.  
Notice that the introduction of a complex energy is equivalent to assuming 
a quasi-steady outgoing flow for the decay. 
The decay
width can then be computed by diagonalizing the non-Hermitian
$\vH$ to find the imaginary part of the eigenenergy of
the appropriate eigenfunction.  However, the interpretation
is complicated by the strong dependence of the decay width on the energy
difference $E_d - E_g = \Delta - i \Gamma_d/2$.  
As
was shown in Ref. \cite{HB20}, it is helpful
to make an approximate reduction of $\vH$  to an effective
2$\times$2 Hamiltonian matrix 
\begin{equation}
\vH_{\rm eff} = 
\left(
\matrix{
E_{g} & v_{\rm eff} \cr
v_{\rm eff} & E_g+ \Delta -i\Gamma_d/2 \cr}
\right),
\label{2x2}
\end{equation}
where  
\begin{equation}
v_{\rm eff}=\vec{v}_g
\left[
E_g \mathds{1} - \vH_b \right]^{-1} \vec v^T_d.
\label{veff}
\end{equation}
Here $\mathds{1}$ is the unit matrix. This formulation also has
the advantage that one avoids the computational issues associated with
diagonalizing large non-Hermitian matrices.  
The coupling matrix element $\ve$ is generally small enough to 
be treated perturbatively, in which case the decay width $\Gamma_f$ is given by
\be
\Gamma_f \approx \frac{\Gamma_d\ve^2}{\Delta^2+\Gamma_d^2/4}.
\ee
Thus the accuracy of approximations to
$\left[
E_g \mathds{1s} - \vH_b \right]^{-1}$
can be assessed from comparing their  derived $v_{\rm eff}$ values.
The main calculational problem is inverting the large
matrix in Eq. (\ref{veff}).

\subsection{Large configuration spaces}

The calculational problem of evaluating Eq. (\ref{veff}) can be 
simplified if the configurations in $\vH_b$ 
can be ordered
by some attribute such as the degree of elongation or the changes in
orbital occupation numbers. 
Such configurations may be constructed e.g., in the Hartree-Fock 
approximation with a constraint on shape degrees of freedom \cite{BY19,BYR18,BYR19}.
The 
configurations that are well separated in the ordered list are not 
directly connected by 
the Hamiltonian, and the matrix can therefore be considered
to be block-tridiagonal,
\begin{equation}
\vH_b\approx \left(
\matrix{
\vH_{1} & \vV_{1} & 0 & 0 & \cdots  \cr
\vV_{1}^T & \vH_{2} & \vV_{2} & 0 & \cdots \cr
0 &  \vV_{2}^T & \vH_{3} & \vV_{3} & \cdots \cr
0 & 0 & \vV_{3}^T & \vH_{4} & \ddots \cr
\vdots & \vdots & \vdots & \ddots & \ddots  \cr}
\right).
\label{tridiagonal}
\end{equation}
with $N_b$ blocks.

Similar Hamiltonians also occur in the theory
of electron transport in carbon nanotubes \cite{Nardelli1999} and
other structures \cite{Reuter11}, and the same calculational techniques
can be applied here.  As is well known, block
tridiagonal matrices can be inverted by Gaussian elimination 
operating on the blocks rather the individual elements of the
matrix (see Appendix).  The block form of Gaussian elimination still requires
inverting the diagonal block matrices, but their dimensions
are much smaller when there are many blocks in $\vH_b$.  
Further speedups are possible if the matrix has a block Toeplitz
form\footnote{ A Toeplitz matrix has all blocks on the same
diagonal equal, i.e.  $\vH_n = \vH_T$ and $\vV_n = \vV_T$ independent
of $n$.}
 \cite{Reuter11,Reuter12,hua97}.  
Later we will  derive an asymptotic expression for
the suppression of the tunneling rate as a function of the
number of blocks in a block Toeplitz $\vH_b$ of special form.

\subsection{Adiabatic and maximum-coupling approximations}
A very common approximation used in the WKB approach is the
adiabatic treatment of the wave function under the barrier.  The
equivalent approximation in the CI approach is obtained by
projecting the block matrices $\vH_n$  
onto the local ground state.  
To this end, we first diagonalize the $\vH_n$ 
to
find the local ground states, 
\begin{equation}
\vH_n \vec\psi^{({\rm ad})}_n = E_n^{({\rm ad})}\vec\psi^{({\rm
ad})}_n. 
\label{adiabatic}
\end{equation}
Then we use the projection operators $P_n = |\vec\psi^{({\rm ad})}_n\rangle
\langle\vec\psi^{({\rm ad})}_n|$
to reduce $\vH$ to the $(N_b+2)$-dimensional matrix
\begin{equation}
\vH_{\rm ad}=\left(
\matrix{
E_g & {v}_g & 0 & \cdots  & 0 \cr
{v}_g & E_1^{({\rm ad})} & {v}^{\rm ad}_1 & \cdots & 0\cr
\vdots & \ddots & \ddots & \ddots & \vdots \cr
0 & 0 & {v}^{\rm ad}_{N_b-1} & E_{N_b}^{({\rm ad})} & {v}_d \cr
0 & 0 & 0 & {v}_d & E_d \cr}
\right),
\label{adiabaticH}
\end{equation}
where  $N_b$ is the number of blocks, and  
$v^{\rm ad}_n=\langle \vec\psi^{({\rm ad})}_n|\vV_n|\vec\psi^{({\rm
ad})}_{n+1}\rangle$, $v_{g}=\langle \phi_{g}|\vec{v}_g
|\vec\psi^{({\rm ad})}_n\rangle$ and similarly for $v_d$.

As mentioned earlier, realistic calculations in the collective-coordinate approach
have shown 
that the barrier penetration integral can be increased substantially by
using wave functions that have stronger
pairing condensates \cite{Robledo14-2}.  The least-action
approach chooses a pairing
condensate having a strength that minimizes the action integral 
along the tunneling path.  
It was suggested in Ref. \cite{HB20} that the least-action
treatment of the WKB penetrability could be simulated in the
CI approach in a similar way, replacing the local ground-state  wave 
functions used to construct $\vH_{\rm ad}$  by wave functions that
were more strongly paired.  
The maximum-coupling approximation is to choose the pairing strength
which maximizes the derived $\ve$.

\section{Application to the schematic model}
In this section we test various approximations with the schematic
pairing-plus-quadrupole
model introduced earlier \cite{Bertsch20,HB20}.  
For completeness, 
we first summarize details of the model as presented in those
publications.
The Fock-space Hamiltonian is defined as
\begin{equation}
\hat H=\sum_{k=0}^{N_{\rm orb}-1}\epsilon_k\hat{n}_k
+v_Q\hat{Q}\hat{Q}-G\sum_{k\neq k'}\hat{P}^\dagger_k\hat{P}_{k'}
\label{hamiltonian}
\end{equation}
where $\hat n_k=a_k^\dagger a_k+a_{\bar{k}}^\dagger a_{\bar{k}}$ is 
the number operator for orbital $k$, $\hat{Q}=\sum_kq_k\hat{n}_k$ represents 
the quadrupole moment, and $\hat{P}_k=a_k^\dagger a_{\bar{k}}^\dagger$ is the  
pair creation operator.  

The specific Hamiltonian treated numerically acts in
a space of 6 doubly degenerate
orbitals, $k=0,\cdots,5$, containing 6 paired particles, which
we call the $(6,6)$ model.  The single-particle energies are
given by

\begin{equation}
\epsilon_k=(k~{\rm mod}~(N_{\rm orb}/2))\epsilon_0,
\label{epsk}
\end{equation}
where $\epsilon_0$ is the single-particle level spacing.  The quadrupole
moments  are set to   $q_k = (-1,-1,-1,1,1,1)$ for the six orbitals.
The parameter $\epsilon_0$ sets the energy scale for the Hamiltonian.
In terms of it, the other  
numerical parameters are chosen in the following way. 
We estimate $\epsilon_0$ at $Q=0$ using the Fermi gas approximation to
obtain $\epsilon_0\sim 2/3$ MeV as a typical value for the actinide nuclei. 
The fission barrier height in the actinide region is 
$B\sim$ 6 MeV \cite{Vandenbosch73}, which is realized in the present (6,6) model 
with $v_Q=-13\epsilon_0/32$. 
The strength of the pairing gap is chosen to be $G=0.563\epsilon_0$ in order to 
reproduce $\Delta=1$ MeV in the BCS approximation with the single-particle
spectrum of the $(6,6)$ model.
These parameters are slightly different from those in Ref. \cite{HB20}, 
but we have confirmed that the conclusions in Ref.\cite{HB20} remain 
the same with the new parameter set.

The active model space includes only seniority-zero states, 
that is, configurations with 3 pairs in the 6 orbitals.  
There are four sets of
configurations in the active space, distinguished by expectation
values of the
$\hat{Q}$ operator, $Q=-6, -2, 2$, and 6.  As indicated in Fig. 
1, there is one configuration at $Q=-6$, the model ground state,
and one configuration at $Q=+6$, which we take as the doorway state
to fission. These two configurations are separated by a barrier
formed by two blocks of configurations at $Q=\pm 2$.  The resulting
Hamiltonian has the form
\begin{equation}
\vH=\left(
\matrix{
E_{g} & \vec{v}_g^T & 0 & 0\cr
\vec{v}_g  & \vH_9 & \vV_9 & 0 \cr
0 & \vV_9 & \vH_9 & \vec{v}_g \cr
0 & 0 & \vec{v}_g^T & E_g  \cr}
\right).
\label{hamiltonian2}
\end{equation}
Here  $E_g$ is the energy of the ground
state energy and $\vec{v}_g$ represents the coupling of the
ground state and doorway to configurations in the barrier region.
The blocks $\vH_9$ and $\vV_9$ are 9$\times$9 dimensional matrices.
As in the previous section, we reduce the Hamiltonian 
to an effective 2$\times$2 matrix
of the form  Eq. (\ref{2x2}).  Its  coupling matrix 
element is given by 
\begin{equation}
v_{\rm eff}=(\vec{v}_g~0)\left[E_g\mathds{1}-
\left(
\matrix{
\vH_9 & \vV_9 \cr
\vV_9 & \vH_9 \cr}
\right)\right]^{-1}
\left(
\matrix{
0 \cr
\vec{v}_g\cr}
\right).
\label{veffa}
\end{equation}
The numerically exact  value of $\ve$ for the assigned model parameters 
is given on the first line in Table I.
\begin{table}
\caption{Comparison of the adiabatic and maximum-coupling
(MC) approximations to the exact coupling matrix element in
the $(6,6)$ schematic model, as calculated by 
Eq. (\ref{veffa}).  
}
\begin{tabular}{c|c|c}
\hline
\hline
Model  & $v_{\rm eff}$ & 
$\left(v_{\rm eff}/v_{\rm eff}({\rm exact})\right)$ \\
\hline
exact &  0.0650  &    \\
adiabatic &  0.0221 & 0.34  \\
MC   & 0.0653  & 1.00 \\
\hline
\hline
\end{tabular}
\end{table}
\begin{figure} [tb]
\includegraphics[width=0.9\columnwidth,clip]{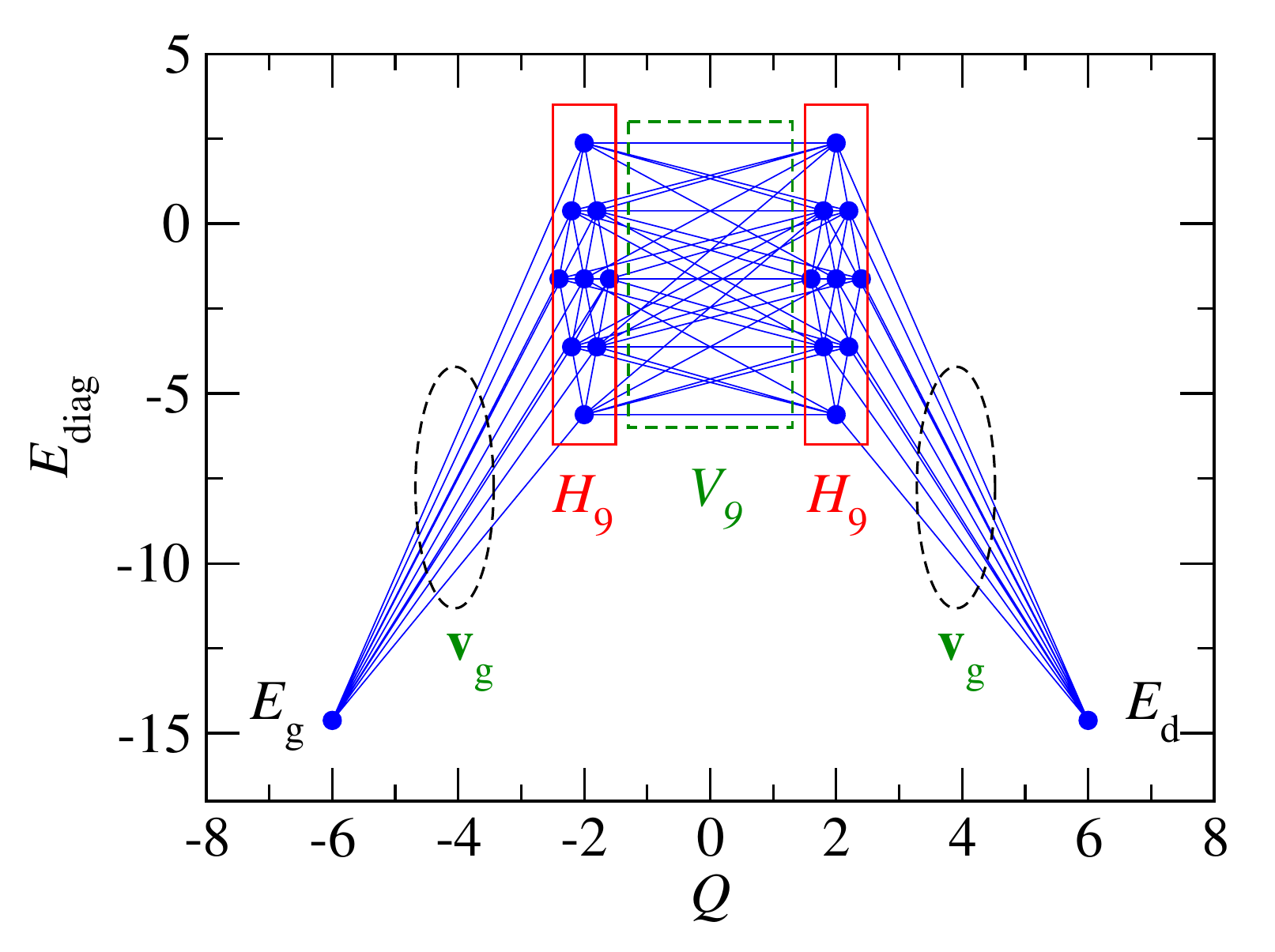}
\caption{
\label{fig1}
The model space for $(N_{\rm orb},N_{p})=(6,6)$ 
in the schematic discrete-basis model. 
The filled circles denote 
the diagonal energies, while 
the lines show the non-zero Hamiltonian matrix elements connecting
configurations. 
The configurations contained in the interior
diagonal blocks are enclosed in solid-line rectangles.  The off-diagonal
block $\vV_9$ is indicated by the matrix elements passing through the
dashed-line rectangle.  Finally, the coupling vectors $\vec{v}_g$ and
$\vec{v}_d$ are indicated by the matrix elements passing through the dashed-line
ellipses.
}
\end{figure}

\subsection{Adiabatic approximation}
The adiabatic approximation reduces the 20$\times$20 dimensional Hamiltonian to a
4$\times$4 matrix.  The effective coupling in the further
reduction to Eq. (\ref{2x2})  is
\begin{equation}
v_{\rm eff}^{({\rm ad})}=\frac{v_{\rm ad}v_g^2}{\left(E_b^{({\rm ad})}\right)^2-v_{\rm ad}^2},
\end{equation}
where $E_b^{({\rm ad})}=E_0^{({\rm ad})}-E_g$ is the adiabatic barrier
height.   The numerical value is shown on the second line of Table I. 
As we showed in Ref. \cite{HB20}, 
the adiabatic approximation considerably underestimates $v_{\rm eff}$ with 
the suppression factor 
$S\equiv (v_{\rm eff}^{({\rm ad})}/v_{\rm eff})^2$ of 0.116 with the present 
parameter set.  A further calculation with a larger model spaces
containing 3-4 internal blocks showed an even larger
suppression factors.

\subsection{Maximum coupling approximation}
Guided by the findings in the collective-coordinate  approach, we introduce 
the pairing condensate as a
dynamical variable.
\begin{figure} [tb]
\includegraphics[width=0.9\columnwidth,clip]{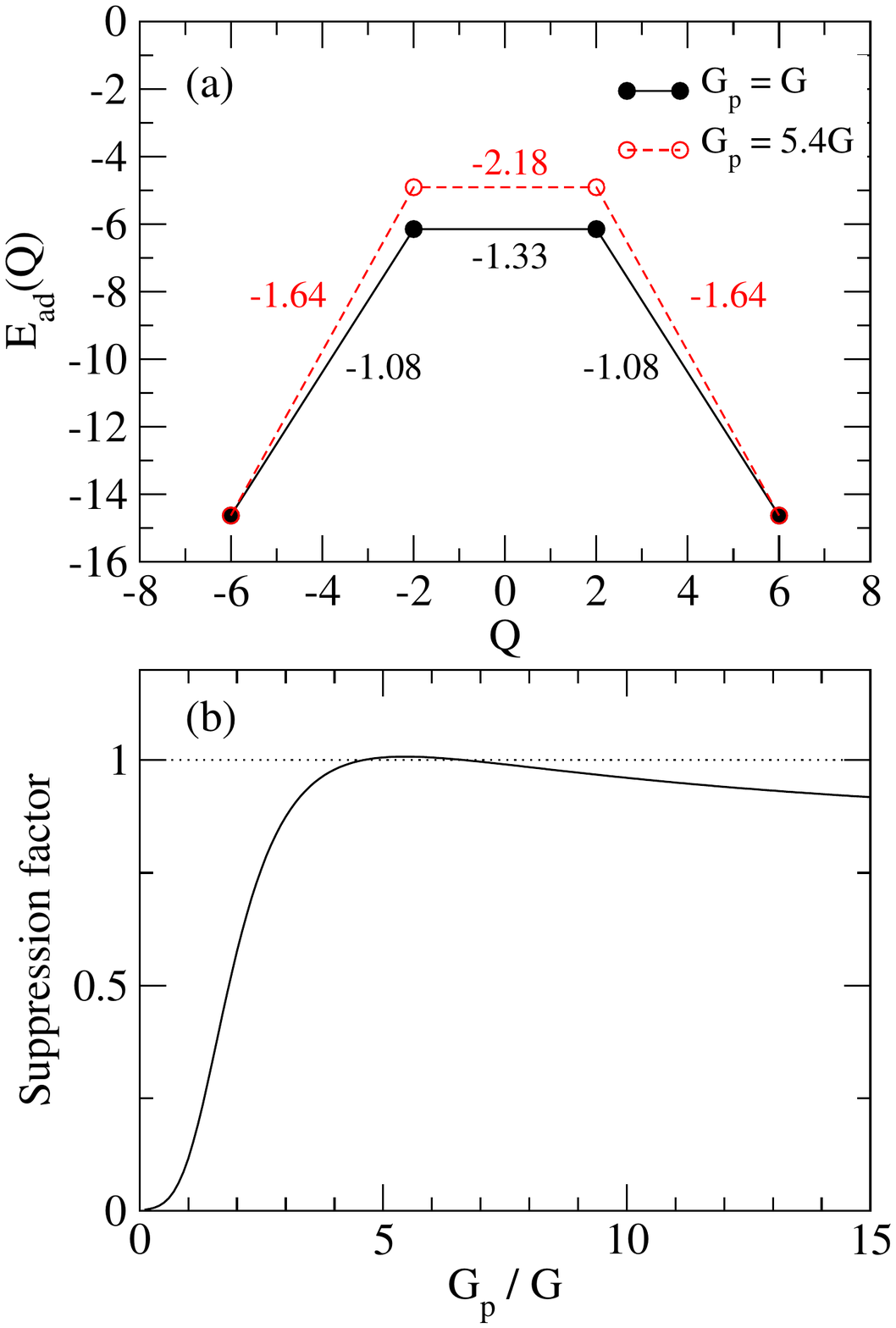}
\caption{\label{local_S}
Panel a): comparison of energies and Hamiltonian matrix elements
in the adiabatic
approximation (the solid line with
filled circles) to those obtained in the maximum-coupling approximation.
(the dashed line with open circles).
The numbers denote coupling strengths connecting the neighboring
adiabatic states.
Panel b): 
the suppression factor,
$S(G_p)\equiv (v_{\rm eff}^{({\rm ad})}(G_p)/v_{\rm eff})^2$,
as a function of $G_p/G$.
}
\end{figure}
In the present approach, this
corresponds to changing the ground state wave function
in Eq. (\ref{adiabatic}) to the lowest state of the block Hamiltonian
having $G$ changed to $G_p$,
that is, the solution to the eigenvalue equation
\begin{equation}
 H_9(G_p)|\phi^{({\rm ad})}_0(G_p)\rangle = E_0^{({\rm ad})}(G_p)
 |\phi^{({\rm ad})}_0(G_p)\rangle. 
\label{Gp}
\end{equation}
The model Hamiltonian with the original $G$ is then diagonalized
with the basis defined by
$\phi^{({\rm ad})}_0(G_p)$; the optimum pairing strength is
the one giving the largest $\ve$.
We call this the maximum-coupling approximation. 
The various matrix elements in the 4$\times$4 reduction are shown in 
the upper panel of Fig. 2. The solid line with 
filled circles show the interaction matrix element $v_g,v_d$ and
$v^{\rm ad}$ in the adiabatic approximation,
that is, with $G_p=G$. 
The dashed line with open circles denotes the same quantities obtained
with $G_p=5.4 G$. 
Note that the fission barrier is higher,
since the energy is not minimized. 
The increase of the barrier height is more than compensated for by the
increase in $v_g$ and $v_{\rm ad}$.
Fig. 2(b) shows the suppression factor, 
$S(G_p)\equiv (v_{\rm eff}^{({\rm ad})}(G_p)/v_{\rm eff})^2$, as a function
of $G_p/G$. One can see that the suppression factor has a  maximum
near  $G_p/G=5.4$, at which  point the exact $v_{\rm eff}$ 
is well reproduced. 

\subsection{Block Toeplitz modeling}

Realistic CI modeling of actinide nuclei might require of
the order of
$\sim$$20$ configuration sets to describe the physical
barrier region \cite[Fig.8]{BYR19}.
The $(6,6)$ model with its two sets is too
oversimplified to simulate the behavior of
long chains of intermediate configuration.  However, we
simulate the chains 
by replicating the $\vH_9$ and $\vV_9$ blocks along the 
diagonal and subdiagonals.  The Hamiltonian is then characterized by
the number $N_b$ of $\vH_9$ blocks along the main diagonal.  Such
matrices are known as block Toeplitz matrices; 
as mentioned earlier there has been much effort in other
fields to find efficient algorithms to invert them.  It turns
out that for the $(6,6)$ Hamiltonian one can extract the
asymptotic dependence on $N_b$ from the properties of the
eigenvalues of a matrix having 
the same dimension as $\vH_9$.

\begin{table}
\caption{
The effective coupling strength $v_{\rm eff}$ for the
block tridiagonal matrix, Eq. (\ref{tridiagonal}), as
a function of the number of blocks $N_b$.
$\vH_9$ and $\vV_9$ are taken from the (6,6) model. 
The ratio to the
effective strength for $N_b-1$ is also shown. The last
row shows the ratio $e^{-\alpha}$ as computed from 
Eq. (\ref{acosh}). }
\begin{tabular}{c|l|c}
\hline
\hline
$N_b$ & $v_{\rm eff}(N)$ & $v_{\rm eff}(N_b)/v_{\rm eff}(N_b-1)$ \\
\hline
1  & 0.277 & -- \\
2  & 0.0650 & 0.2345 \\
3  & 0.0154 & 0.2372 \\
4  & 0.00367 & 0.2380 \\
5  & 0.000875 & 0.2384 \\
6  & 0.000209 & 0.2385 \\
7  & 0.0000498 & 0.2386 \\
8  & 0.0000119 & 0.2386 \\
\hline
\multicolumn{2}{c|}{Eq. (\ref{acosh})} & 0.2386 \\
\hline
\hline
\end{tabular}
\end{table}
First, we show in Table II the effective coupling strength $v_{\rm eff}$
as a function of $N_b$. See also Fig. 3. 
As is expected, $v_{\rm eff}$ decreases as $N_b$ increases.
The Table and the figure 
also show the ratio of $v_{\rm eff}(N)$ to $v_{\rm eff}(N-1)$. 
One can see that
the ratio goes asymptotically
to a constant $C$ which we call the $N_b$-scaling factor,
\be
C = \lim_{N\rightarrow \infty} 
 v_{\rm eff}(N)/v_{\rm eff}(N-1) = 0.2386. 
\ee
\begin{figure} [tb]
\includegraphics[width=0.9\columnwidth,clip]{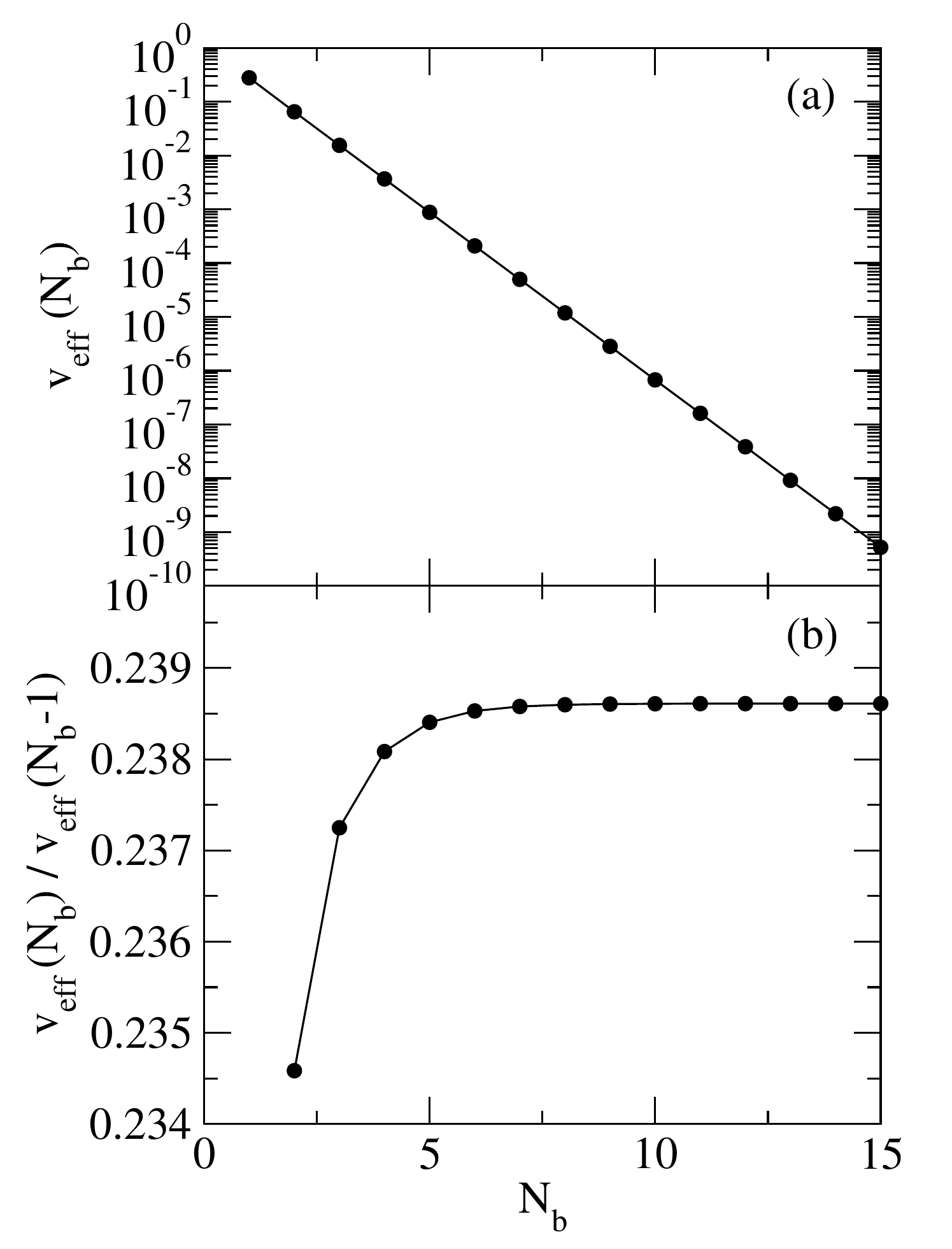}
\caption{
The effective coupling strength $v_{\rm eff}$ 
(the upper panel) 
and 
the ratio to the
effective strength for $N_b-1$ 
(the lower panel) as a function of the number of blocks $N_b$. }
\end{figure}

A formula for $C$ can be derived as follows, assuming 
a Toeplitz structure of $H_b$ with symmetric $V_T$.
We start with the equation 
for the wave function
\begin{equation}
\vH
\left(\matrix{
  \phi_g \cr
  \vec{\psi} \cr
  \phi_d \cr}
\right)
= E_g
\left(\matrix{
 \phi_g \cr
 \vec{\psi} \cr
 \phi_d \cr}
\right),
\end{equation}
valid when $E_d=E_g$.  
Writing the $n$-th block component of the interior wave function
as $\vec\psi_n$,  each block row has the form
\begin{equation}
\vV_T\vec\psi_{n-1}+\vH_T\vec\psi_n+\vV_T\vec\psi_{n+1}=E_g\vec\psi_n.
\label{tridiagonal-eigen}
\end{equation}
Note that $V_9^T=V_9\equiv V_T$ in this equation. 
Eq. (\ref{tridiagonal-eigen}) is invariant under translation of 
the block indices $n$.
That implies that the wave function can be expressed as a sum over 
amplitudes that vary from block to block as $\vec\psi_{n+1}^\lambda = 
C_\lambda \vec\psi_{n}^\lambda$
where $C_\lambda$ is a constant.
For tunneling under a barrier,  $C_\lambda$ are all real and can be
written as 
$C_\lambda = e^{\pm\alpha_\lambda}$.
Substituting in Eq.
(\ref{tridiagonal-eigen}) one obtains
\begin{equation}
(e^{\alpha_\lambda}+e^{-{\alpha_\lambda}})\vV_T\vec\psi_n
+(\vH_T-E_g\mathds{1})\vec\psi_n=0.
\label{tridiagonal-eigen3}
\end{equation}
This is equivalent to the eigenvalue equation 
\begin{equation}
\left(-2(\vH_T-E_g\unit)^{-1}\vV_T\right)\vec\psi_n=\lambda \vec\psi_n 
\label{tridiagonal-eigen2}
\end{equation}
with $\alpha_\lambda$ related to the eigenvalue by 
\be
\cosh(\alpha_\lambda) = \lambda^{-1}. 
\ee
The wave function will be decaying going from the ground state toward
the doorway configuration so we may assume that the asymptotic behavior
is 
\begin{equation}
\vec\psi_{n+1}=e^{-\alpha_\lambda}\vec\psi_n
\label{exponential}
\end{equation}
with $\alpha_\lambda > 0$.
Thus  $v_{\rm eff}$ decreases when a block is added by
\be
C = e^{-\alpha_\lambda} = e^{-\cosh^{-1}(|\lambda|)}
\label{acosh}
\ee
where $\lambda$ is the eigenvalue of Eq. 
(\ref{tridiagonal-eigen2}) having the largest absolute value. 
Note that eigenvalues with $|\lambda| > 1$ correspond to
undamped propagation modes.  If such eigenvalues are present,
the physical conditions for barrier penetration are violated.

The last line in Table II shows 
the $N_b$-scaling factor $C$ derived from Eq.
(\ref{acosh}) for the $(6,6)$ model.  The agreement with the
observed reduction factor is excellent.  

In a more general case with $V\neq V^T$, we have found two ways to
compute $C$.
One is somewhat parallel to the above argument, but starting from the
block row equation
\begin{equation}
(\vV_T)^T\vec\psi_{n-1}+\vH_T\vec\psi_n+\vV_T\vec\psi_{n+1}=E_g\vec\psi_n,
\label{tridiagonal-eigen4}
\end{equation}
rather than Eq. (\ref{tridiagonal-eigen}).  The other method is based
on partial Gaussian elimination and is presented in the Appendix.

For the first method, we assume
$\vec\psi_{n\pm1}=e^{\mp \alpha}\vec\psi_n$ as in Eq. 
(\ref{tridiagonal-eigen3}).  Substituting in  Eq. 
(\ref{tridiagonal-eigen4}), one obtains 
\begin{equation}
\left[(\vV_T)^Te^\alpha+\vH_T-E_g\mathds{1} 
+\vV_Te^{-\alpha}\right]\vec\psi_{n}
\equiv M(\alpha)\vec\psi_{n}=0. 
\end{equation}
This equation is satisfied when the determinant of $M(\alpha)$ is zero. 
This can be solved numerically for $\alpha$, from which the $N_b$-scaling factor 
is evaluated as $C = e^{-\alpha}$. 

We have verified that this method 
yields the same value of the $N_b$-scaling factor as that 
obtained with Eq. (\ref{acosh}) when $\vV=\vV^T=\vV_9$. Moreover, we have also 
confirmed that this method leads to the exact scaling factor when some of the 
components in $\vV_9$ is set to be zero so that $\vV_9\neq (\vV_9)^T$.

\section{LINK TO LEAST ACTION APPROACH}

\subsection{Derived action in the CI framework}
\begin{figure} [tb]
\includegraphics[width=0.9\columnwidth,clip]{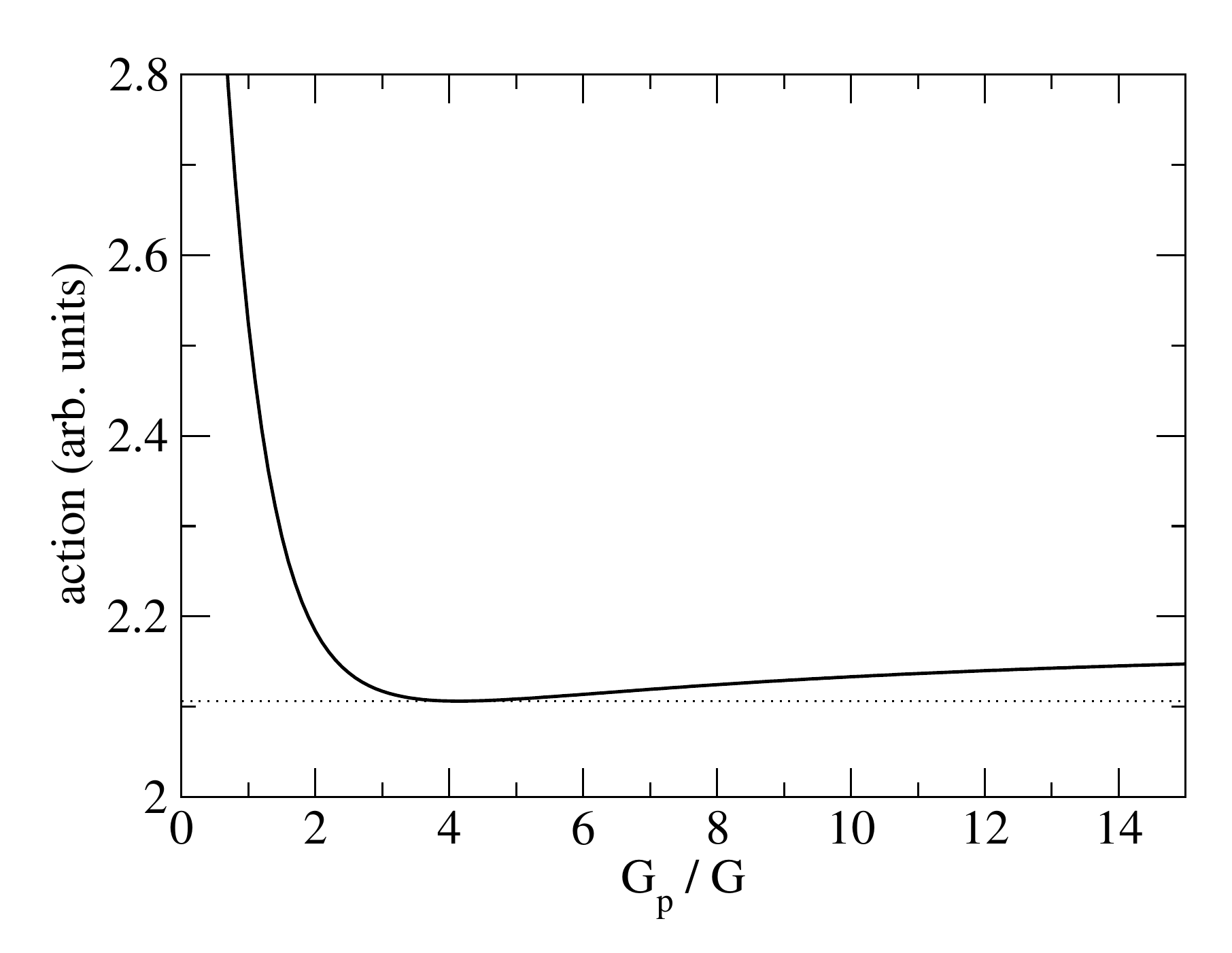}
\caption{
The local action $S\propto \sqrt{B/|v|}$ as a function of $G_p/G$.
Here, the barrier height $B$ and the coupling strength $v$ are estimated
with the lowest eigenstate of the intermediate Hamiltonian $H_9$
with a modification of the strength of the pairing interaction from $G$
to $G_p$. }
\end{figure}

The action integral in the collective-coordinate  approach is given by
\be
S = \int \left[ \frac{2I(x)}{\hbar^2}(V(x)-E_g) \right]^{1/2} dx;
\ee
where $I$ is the inertia of the system,  often calculated in the
Gaussian overlap approximation \cite[Sect. 10.7.4]{R&S}. The variable 
$x$ is a shape degree of freedom associated with the fission path,
and
$V(x)$ is the energy of
the GCM configuration at the point $x$. 
We want to minimize $S$ given the $\vH_b$ matrix of block Toeplitz
form.  We first reduce the matrix to the form 
\begin{equation}
H_b=\left(
\matrix{
 e & v & 0 & 0  \cdots  & 0 \cr
 v & e & v & 0  \cdots & 0 \cr
 0 &  v & e & v  \cdots & 0 \cr
 \vdots & \vdots & \ddots & \ddots & \ddots  \cr
 0 & 0 & 0 & v & e \cr}
\right),
\label{tridiagonal2}
\end{equation}
with the aid of a one-dimensional projection operator acting on
the $H_T$ and $V_T$ matrices.  
The equation which corresponds to Eq. (\ref{tridiagonal-eigen})
then reads
\begin{equation}
v\psi_{n-1}+e\psi_n+v\psi_{n+1}=E_g\psi_n.  
\end{equation}
This is to be compared with  a Schr\"odinger equation for a
collective Hamiltonian,
\begin{equation}
\left(-\frac{\hbar^2}{2I}\frac{d^2}{dx^2}+V(x)\right)\phi(x)=E\psi(x).
\label{collective}
\end{equation}
Discretizing the differential operator as 
\begin{equation}
\frac{d^2}{dx^2}\phi(x)\sim \frac{1}{(\Delta x)^2}
\left[\psi(x_{n-1})-2\psi(x_n)
+\psi(x_{n+1})\right], 
\label{discretization}
\end{equation}
where $\Delta x$ is a mesh spacing of the variable $x$,
the inertia parameter $I$ and the collective potential $V(x)$ can be
read off as \cite{Barranco90}
\begin{eqnarray}
I&=&-\frac{\hbar^2}{2v}\frac{1}{(\Delta x)^2}, \\
V(x)&=&e-2v. 
\end{eqnarray}
Since the collective potential has a constant shift $-2v$ at all
the points of $x$, we expect that the energy $E$ in Eq. (\ref{collective})
is approximately given by $E=E_g-2v$.
Introducing exponentially decaying wave functions in Eq. (\ref{collective}), 
\begin{equation}
\psi(x_{n-1})=e^{S\Delta x}\psi(x_n),~~~\psi(x_{n+1})=e^{-S\Delta x}
\psi(x_n),
\end{equation}
one obtains
\begin{equation}
\cosh(S\Delta x)=(IB(\Delta x)^2+1),
\label{cosh}
\end{equation}
with $B\equiv e-E_g$.
To be consistent with the discretization in Eq. (\ref{discretization}),
we expand the lefthand side of this equation to obtain
\begin{equation}
\cosh(S\Delta x)\sim 1+\frac{S(\Delta x)^2}{2}=(IB(\Delta x)^2+1),
\end{equation}
from which the local action $S$ reads
\begin{equation}
  S\sim \sqrt{2BI}
\end{equation}
and its increment from block to block is
\be
S\Delta x/\hbar = \sqrt{
\left(\frac{B}{-v}\right)}.  
\label{Sdel}
\ee
Figure \ref{local_S}  shows the action as a function of $G_p/G$,
where $B$ and $v$ 
are estimated with
$\phi^{({\rm ad})}_0(G_p)$ in Eq. (\ref{Gp}).
As in the maximum-coupling approximation discussed in Sec. II B,
both $B$ and $|v|$ increase as a function of $G_p$, but the ratio has
a minimum at some large value of $G_p$.  With the parameter
set which we employ for the (6,6) model, the minimum of the action is
found at $G_p/G=4.1$. This is close to the value
$G_p/G=5.4$ we obtained 
in the maximum-coupling approximation.  In fact  
the suppression factor is not sensitive to the value of $G_p$ around the 
optimum value as may be seen in Fig. 2(b). 

\subsection{Connection to the ``number fluctuation''}

In Refs. \cite{Robledo14,Robledo18}, 
the least-action formalism is applied by treating the  number fluctuation
as a collective variable in  Hartree-Fock-Bogoliubov (HFB) wave functions.  
In terms of the HFB canonical variables $u_k,v_k$,
it is given by \cite[Eq. 6.44]{R&S} 
\be
\Delta N^2= 4\sum_{k>0} v_k^2u_k^2=4\sum_{k>0} v_k^2(1-v_k^2), 
\label{numberfluctuation}
\ee
A related measure of the pairing strength is the abnormal density $\kappa$,
defined for a $P^\dagger P$ pairing interaction as 
\be
\kappa=\sum_{k>0} u_kv_k=\sum_{k>0} v_k\sqrt{1-v_k^2}. 
\label{kappa}
\ee
Since the CI approach is a 
number-conserving framework and has no number fluctuation, one cannot
make a direct comparison to HFB pairing. 
But one can still make a connection through the
orbital occupation numbers $n_k$ in the CI local ground state,
\begin{equation}
n_k=\langle \phi^{({\rm ad})}_0(G_p)|a_k^\dagger a_k|\phi^{({\rm ad})}_0(G_p)\rangle. 
\label{nk}
\end{equation}
We make the identification $v_k^2 = n_k$ to relate the CI wave function to
the HFB quantities in Eq. (\ref{numberfluctuation}) or (\ref{kappa}).
The result comes out to $\kappa = 1.46$ and $ \Delta N^2 = 1.50$ for the
adiabatic wave function ($G_p=G$).  Fig. 5 shows how the quantities 
relate to the action change with the two measures of pairing strength.
\begin{figure} [tb]
\includegraphics[width=0.9\columnwidth,clip]{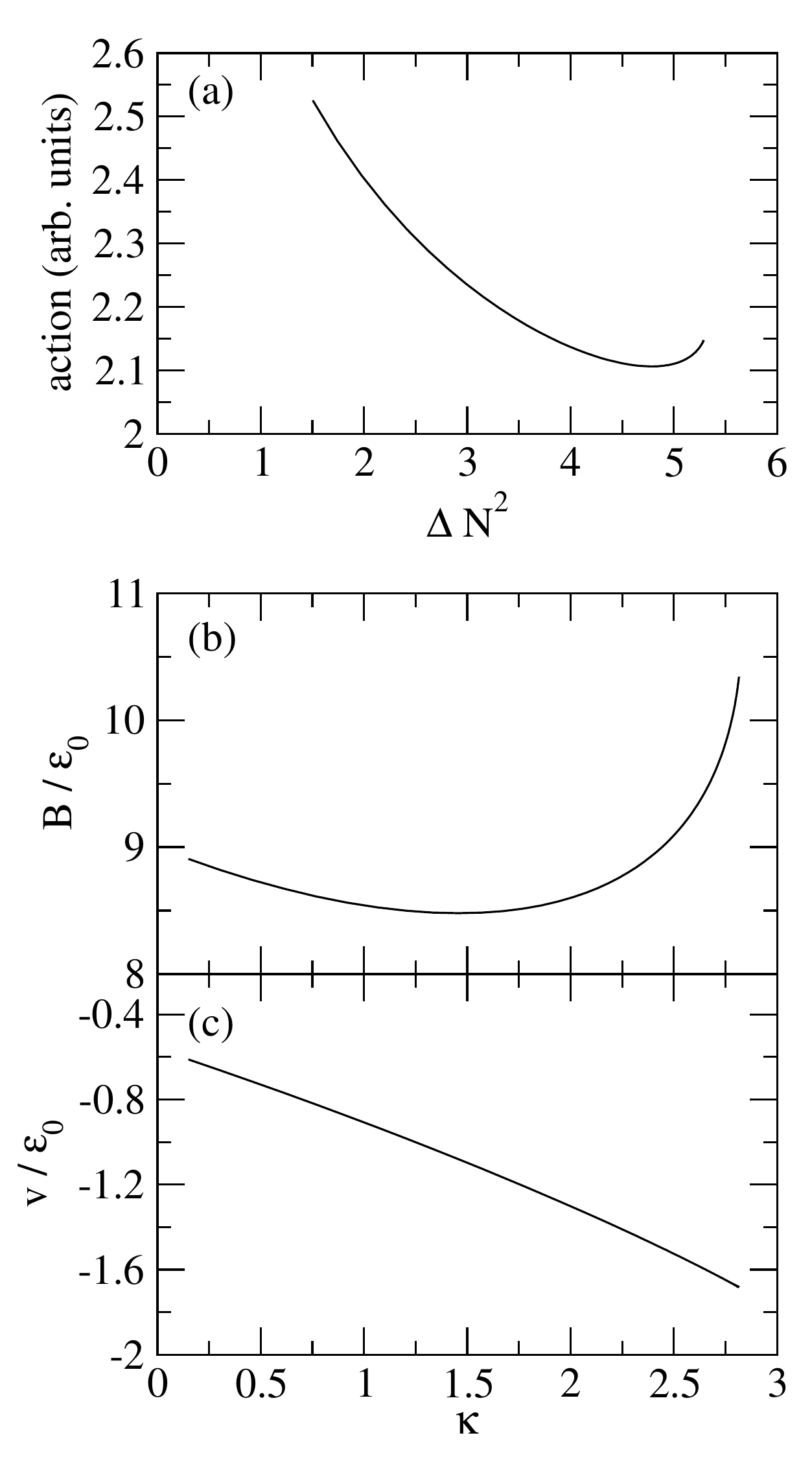}
\caption{
(The upper panel) The same as Fig. 4, but as a function of the 
number fluctuation $\Delta N^2$ defined by Eq. (\ref{numberfluctuation}). 
(The middle and the bottom panels) The barrier height $B$ and the 
interaction $v$ as a function of the abnormal density defined by Eq. (\ref{kappa}). 
}
\end{figure}

Fig. 5(a) shows the action increment Eq. (\ref{Sdel}) as a function of $\Delta N^2$. 
It decreases to 4.7 at $G_p=4.1 G$ where it has a 
minimum. The ratio is 3.13, which is close to 
the ratio $\sim 3.17$ for $^{234}$U shown in Ref. \cite{Robledo14}. 
Figs. 5(b) and 5(c) show the barrier height $B$ and the interaction $v$ 
as a function of the pair density $\kappa$.  These plots are shown to 
compare with the known  small-amplitude behavior in BCS theory.  There the
leading behavior is $ v \sim \kappa^2$ and $B \sim B^{\rm ad} + (\kappa -\kappa^{\rm
ad})^2$ for B \cite{Moretto74}.  A fit to these functional forms would be quite poor.
But it might be that the validity of the quadratic formulas need a much larger
orbital space than we have in the $(6,6)$ modeling. 

\section{Summary}

Using a schematic configuration-interaction
model for spontaneous fission, 
we have investigated two approximations that go beyond the adiabatic
approximation. To this end, we have constructed a single configuration
at each shape by increasing the strength of the pairing interaction. This
was motivated by the fact the pairing fluctuation makes an important
degree of freedom in describing spontaneous fission based on the density
functional approach. An increase of the pairing interaction results in 
an increase of both the barrier height and the coupling strength
between 
neighboring configurations. 
In the first approximation, we investigated the maximum
coupling approximation, in which the optimum value of the modified pairing
strength is determined so that the effective coupling strength $v_{\rm eff}$
between the ground state and the fission doorway state is maximized.
In this
connection, we have investigated a chain of the interior matrix and have found that
the effective coupling strength has a scaling property as a function of
the number of blocks. We have shown that the scaling property can be
understood by a simple eigenvalue equation if one assumes exponentially
decaying wave functions for the fission degree of freedom.

In the second approximation, we have investigated the least action approach,
in which the optimum value of the modified pairing strength is determined to
minimize the action.  This approximation can be derived in the context of the discrete
basis model that we employ in this paper. We have shown that the optimum
value of the modified pairing is close to that in the maximum-coupling
approximation, yielding a reasonable value of the effective coupling strength.
This gives some justification
for treating the barrier penetration by the WKB formula with parameters derived
from the CI approach.

In realistic applications to spontaneous fission, our studies with the
schematic model have indicated that the adiabatic approximation can
considerably underestimate the effective coupling strength, and thus the
decay rate. The two approximations discussed in this paper, that is,
the maximum-coupling approximation and the least action approach,
provide a promising truncation scheme of configurations, by 
taking into account the non-adiabatic effects.
The maximum-coupling approximation can be extended also to the case where
the configurations are not orthogonal to each
other \cite{Nardelli1999,Reuter11}.
The main advantage of the CI approach is that it permits much richer
configuration spaces than can be easily achieved with the GCM or 
pure mean-field dynamics.  However, as mentioned earlier, new computation
tools need to be developed for constructing the spaces and especially
for calculating Hamiltonian matrix elements between arbitrary
configurations.

\section*{Acknowledgments}
We thank J. Dobaczewski, W. Nazarewicz and other participants in the
workshop ``Future of Fission Theory'', York, UK (2019) for discussions
motivating this study.
The work of K.H. was supported by
JSPS KAKENHI Grant Number JP19K03861.

\appendix

\section{Block scaling factor from the Gaussian elimination method}

For a block-tridiagonal matrix (\ref{tridiagonal}), the Greens function 
$G=\left[\vH_b -E_g \mathds{1} \right]^{-1}$ has a form of 
\begin{equation}
\vG=
\left(
\matrix{
  \vG_{11} & \cdots & \vG_{1N} \cr
  \vdots & \ddots & \vdots \cr
  \vG_{N1} & \cdots & \vG_{NN} \cr}
\right).
\end{equation}
Here the subscript $b$ in the number of blocks $N_b$ has been dropped for
clarity. In this notation the effective coupling (\ref{veff}) is given by 
\begin{equation}
v_{\rm eff}=-\vec{v}_g^T \vG_{1N}\vec{v}_d.
\label{veffG1N}
\end{equation}

The matrix $\vG_{1N}$ can be computed by a partial  block-wise Gaussian 
elimination \cite{Nardelli1999,Reuter11,Reuter12,Hod06}.
In this method, 
one first generates matrices $\vA_n$ by iterating 
%
\begin{equation}
\vA_n=\vV_{n-1}^T(\tilde{\vH}_{n-1}-\vA_{n-1})^{-1}\vV_{n-1},~~~~(n=2,3,\cdots,N),
\label{Gaussian1}
\end{equation}
starting from $\vA_1=0$. 
Here, $\tilde{\vH}_n$ is defined as 
$\tilde{\vH}_n\equiv \vH_n - E_g \mathds{1}$. 
We assume all blocks have the dimension, so
$\vA$ matrices are square and of the same dimension.
The $(N,N)$ component of the Greens function $\vG_{NN}$ is given
by 
\be \vG_{NN}=(\tilde{\vH}_N-\vA_N)^{-1}
\ee 
The Greens function component $\vG_{1N}$ in Eq. (\ref{veffG1N}) can be obtained
recursively as
\begin{equation}
\vG_{n,N}=-(\tilde{\vH}_n-\vA_n)^{-1}\vV_n\vG_{n+1,N},~~~(n=N-1,\cdots ,1).
\end{equation}

For our purposes, we do not need the Greens function itself but only
the $N_b$-scaling factor relating  $\vG_{1,N}$ and $\vG_{1,N-1}$.  From
the algebraic structure of the Gaussian elimination quantities it is
easy to show that the relationship can be expressed  
\begin{eqnarray}
\vG_{1N}(N)&=&-\vG_{1,N-1}(N-1)\vV_{N-1}\vG_{NN}(N), \label{Gaussian2} \\
&=&-\vG_{1,N-1}(N-1)\vV_{N-1}(\tilde{\vH}_N-\vA_N)^{-1}. \nonumber \\
\end{eqnarray}
Note that the effective interaction is given by 
$v_{\rm eff}(N)=-\vec{v}_g^T \vG_{1N}(N)\vec{v}_d$ and 
$v_{\rm eff}(N-1)=-\vec{v}_g^T \vG_{1,N-1}(N-1)\vec{v}_d$.

For a block Toeplitz matrix with $\vV_n=\vV_T$ and $\tilde{\vH}_n=\vH_T$, one may 
expect 
that Eq. (\ref{Gaussian1}) produces a sequence that converges to a fixed $\vA
\equiv \vA_\infty$ as $N \rightarrow \infty$.
%
%
In this limit, $\vG_{NN} \rightarrow (\vH_T-\vA_\infty)^{-1}\equiv\vG_\infty$. 
When this is realized, one may also assume that 
the $N_b$-scaling factor $C$ can be computed from 
\begin{equation}
\vG_{1N}(N)=C \vG_{1,N-1}(N-1).
\end{equation}
Substituting this to Eq. (\ref{Gaussian2}), one finds that $C$ may be 
calculated as  $C=\lambda_{\rm max}$, where $\lambda_{\rm max}$ is the 
largest eigenvalue of $-\vV_T\vG_\infty$. 
Note that the iterations in Eq. (\ref{Gaussian1}) need not be extend to
$n=N$ if 
the asymptotic form of $\vA_n$ is approached with fewer iterations.

\begin{table}
\caption{The $N_b$-scaling factor 
corresponding to Table II, but 
obtained with the Gaussian elimination method. 
In this method, the scaling factor is given by 
$\lambda_{\rm max}(n)$, that is, 
the largest eigenvalue of 
$-\vV_T(\vH_T-\vA_n)^{-1}$, where the matrix $\vA_n$ is given by 
Eq. (\ref{Gaussian1}). The table shows this quantity as a function of $n$. 
}

\begin{tabular}{c|c}
\hline
\hline
$n$ & $\lambda_{\rm max}(n)$ \\
\hline
    1  &  0.22576 \\
    2  &  0.23788 \\
    3  &  0.23857 \\
    4  &  0.23861 \\
    5  &  0.23861 \\
    6  &  0.23861 \\
    7  &  0.23861 \\
    8  &  0.23861 \\
    9  &  0.23861 \\
   10  &  0.23861 \\
\hline
\hline
\end{tabular}
\end{table}

Table III shows $\lambda_{\rm max}(n)$, that is, the largest eigenvalue of 
$-\vV_T(\vH_T-\vA_n)^{-1}$, as a function of $n$. One can first see that 
the asymptotic value of this quantity coincides with the $N_b$-scaling factor 
shown in Table II. Secondly, one can see that after a few iterationcs 
$\lambda_{\rm max}(n)$ quickly converges to the asymptotic value.
This implies that $\vG_{NN}(N)=(\vH_T-\vA_N)^{-1}$ can be estimated as 
$\vG_{NN}(N)\sim \vG_{nn}(n)=(\vH_T-\vA_n)^{-1}$ with a much smaller value of $n$ 
compared to the actual $N$.

\end{document}